# Seismotectonics and Slip Behavior of a Submarine Plate Boundary Fault from Seismicity Repeaters and Tomography using a high-resolution earthquake catalog from machine learning

D. Lange[1], Y. Ren[1], and I. Grevemeyer[1]

[1] GEOMAR Helmholtz Centre for Ocean Research Kiel, Wischhofstr. 1-3, 24148 Kiel, Germany

Corresponding author: Dietrich Lange (dlange@geomar.de)

Key Points:

- We created a high-resolution earthquake catalogue for the Blanco Transform Fault System using phase pickers from machine learning.
- Seismicity outlines complex segmentation and lateral changes of seismic behavior, such as pull-apart tectonics and off-axis deformation.
- Repeating events indicate stick-slip at intervals of ~4 years; earthquake tomography reveals high vp/vs ratios related to serpentinization.

**Abstract**

The Blanco transform fault system (BTFS) is highly segmented and represents an evolving transform plate boundary in the Northeast Pacific Ocean. Its seismic behavior was captured with a dense network of 54 ocean-bottom-seismometers operated for one year. We created a high-resolution earthquake catalog based on different machine learning onset pickers, resulting in a high-resolution seismicity catalog with 12.708 events outlining the current deformation and stress release along a major transform fault. Seismicity reveals lateral changes of seismic behavior, indicating seismic and aseismic fault patches or segments, complex along-strike and off-axis deformation, step-overs, and internal faulting within pull-apart basins. Seismicity along simple linear fault strands is localized within 2 km of the seafloor expression of the fault. Repeaters indicate an average of 21 cm slip, exceeding the geological slip rate by ~4 times. Based on the repeater behavior, we suggest that the (overall aseismic) slip is spatially very heterogeneous, consisting of many small seismic patches, each one releasing its seismic slip every 4 years. Along the BTFS, the coupling of the fault is variable and varies between fully locked and fully creeping. Local earthquake tomography shows elevated vp/vs values exceeding 2, suggesting significant serpentinization from seawater entering the transform faults, the oceanic crust and mantle. The study shows how to use modern machine learning pickers on OBS data to provide essential insights into the physics of faulting along major plate boundary faults in time and space, including the partitioning of slip seismic and aseismic faulting with high resolution.

**Plain Language Summary**

The Blanco Transform Fault System is a plate boundary in the Northeast Pacific Ocean that is constantly evolving. We use a dataset of 54 ocean-bottom seismometers to monitor its earthquake activity for a year. By applying machine learning techniques, we located over 12,000 earthquakes, providing a detailed picture of how the fault moves. Some segments experience high earthquake activity, while others move without generating significant seismic activity. One section is locked, meaning stress builds up over time before an earthquake occurs, while others are creeping, meaning they move continuously without major quakes. We identify similar earthquakes, called repeating earthquakes, which allow us to determine the slip along the fault. The repeaters indicate slip exceeding the long-term geological movement. We explain this behavior with repeaters occasionally rupturing the fault with a repeat rate of around 4 years. Additionally, our analysis of seismic velocities (vp/vs ratios) suggests that seawater penetrates deep into the fault system, altering the rocks through serpentinization. Overall, our study demonstrates how modern machine learning tools can improve our understanding of fault behavior, helping us better predict how and when faults might slip in the future.

**1 Introduction**

Transform faults are one of the three major plate boundaries, with a total length of 47,000 km (Bird, 2003). Oceanic transform faults offset the mid-ocean ridges and are responsible for accommodating the movement of tectonic plates as they spread apart during seafloor spreading. Even though seismically active, oceanic transform faults release only a minor fraction of the expected geodetic slip by earthquakes, outlining a major gap in our understanding of how plate motion is accommodated at one of Earth's major plate boundaries. The movement along transform faults is predominately strike-slip, meaning the motion is predominantly horizontal (Wilson, 1965), and the fault trace is visible on the surface of the ocean floor (Sykes, 1967). Because the transform faults are located offshore and the total fault slip cannot be measured, seismicity is often used as a proxy for deformation along transform faults. Therefore, the seismic behavior of most oceanic plate boundaries is characterized by global catalogs with poor completeness offshore (e.g., Aderhold and Abercrombie, 2016) or from land stations (Hellweg et al., 2024). For deployments of ocean-bottom-seismometers (OBS) along transform faults, the number of OBS stations is mostly around 15 stations, with some denser deployments such as the GOFAR transform fault in the Pacific (e.g. Froment et al., 2014) and the Chain transform and BTFS (Kuna et al., 2019; Schlaphorst et al., 2023) with 39 and 54 stations, respectively.

Since the long-term moment release from seismicity is much below the total slip rate estimated (determined from magnetic anomalies and global plate motion models), it is well known that transform faults move mostly aseismic (Brune, 1968), and only ~25% of the total seismic moment accommodated along is released seismically (Boettcher and Jordan, 2004). Repeating earthquakes (Nadeau and Johnson, 1998; Nadeau and McEvilly, 1999; Uchida and Bürgmann, 2019) reveal the transient development of seismic slip in time and space along a fault, but the method was not often used for oceanic transform faults and using OBS. For strong Mw>6.5 events at the Charlie-Gibbs transform fault, Aderhold and Abercrombie (2016) find that the fault moves seismically during the co-seismic and aseismically during the interseismic. The Mendocino transform fault off northern California was found to be 65% aseismic using seismic landstation recordings of repeating earthquakes (Materna et al., 2018). Therefore, the transient total behavior of slip (e.g., seismic plus aseismic slip) remains enigmatic since only limited information on crustal strain from offshore geodetic observation is available (Lange et al., 2019).

Automated phase detection of P and S arrivals on seismograms advanced significantly during the last years using deep learning, but most of the pickers were trained on land data only (Anderson et al., 2025; Mousavi et al., 2020; Münchmeyer et al., 2022; Zhu and Beroza, 2019). However, deep learning approaches are not yet commonly applied to ocean bottom data due to a lack of appropriate training data and models. However, seismicity registered in submarine settings on ocean-bottom-seismometers (OBS) data differs from land stations as it uses an additional hydrophone channel. Bornstein et al. (2024) assembled an annotated database of manually picked local seismicity from 15 OBS deployments, including a smaller dataset for the BTFS (Ren et al., 2023) and trained PickBlueEQtransformer (PickBluePhasenet) based on EQTransformer (Phasenet), respectively. PickBlue significantly outperforms neural networks trained with land stations and models trained without hydrophone data (Bornstein et al., 2024). These machine-learning approaches for OBS data improved the accuracy of phase arrival time estimation. Still, they were not applied to massive OBS datasets and continuous data and benchmarked by comparing seismicity catalogs and different pickers. In general, high-resolution earthquake catalogs are the base for further studies such as local earthquake tomography or repeating event studies.

Here, we created a high-resolution seismicity dataset based on machine learning onset times for the Blanco transform fault system from a dense OBS network (network code X9, Kuna, 2020; Kuna et al., 2019; Nábělek and Braunmiller, 2012). We compare different processing strategies for combining different phase pickers. The resulting seismicity catalogs reveal the faulting of the plate boundary system, showing pull-apart basis, linear features of the BTFS and off-axis faulting with high precision.

To better understand the deformation and structure, we estimate repeaters to estimate the in-situ slip of the BTFS, which reveals the degree of locking for the transform fault. To investigate the degree of serpentinization along the BTFS, we apply a 2D local earthquake tomography method to invert for vp and vp/vs velocity models along the fault system.

**2. Tectonic Setting**

The Blanco transform fault system (BTFS, Figure 1) is located between the Juan de Fuca Ridge and the Gorda Ridge, separating the Pacific and Juan de Fuca plates. The BTFS is ~350 km long and is characterized by a series of segments that offset in a right-lateral motion two north-south trending spreading segments (Braunmiller and Nábělek, 2008; Dziak et al., 1991; Kuna et al., 2019; Ren et al., 2023). Its western and eastern main transform segments are separated by the Cascadia Depression, a short spreading segment occurring roughly at the center of the BTFS (Embley & Wilson, 1992). The Blanco Transform Fault exhibits right-lateral motion with 51 mm/yr (Wolfson-Schwehr and Boettcher, 2019) and hosts seismic activity with moderate-sized events (Braunmiller and Nábělek, 2008a; Cronin and Sverdrup, 2003) of up to magnitude 6.4 in 1928, 1952, 1981 and 1985 (Storchak et al., 2015).

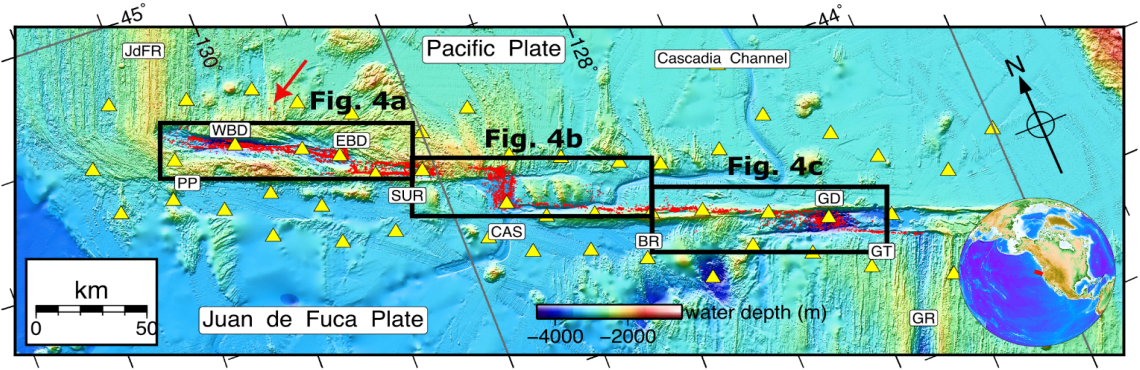

**Figure 1:** Map of the Blanco transform fault system (BTFS) with bathymetry (Ryan et al., 2009) and ocean-bottom-seismometer (OBS) station distribution (yellow triangles). Geologic units are labeled (Embley and Wilson, 1992): BR: Blanco Ridge; CAS: Cascadia Depression; EBD: East Blanco Depression; GD: Gorda Depression; GR: Gorda Ridge; GT: Gorda transform segment; JdFR: Juan de Fuca Ridge; PP: Parks Plateau; SUR: Surveyor Depression; WBD, West Blanco Depression. The red dots indicate the local seismicity from the relative catalog. The red arrow indicates the location of the off-axis seismicity cluster. The inlay shows the location of the BTFS on the globe.

The bathymetry shows no evidence of fossil transform fault traces and, hence, fracture zones around the BTFS before 2 Ma, supporting that there were no transform faults before the initiation of the BTFS. Ren et al. (2023) concluded that the BTFS developed from ridge propagation and the evolution of broad transfer zones instead of resulting from pre-existing transform faults, revealing different slip modes in the eastern and western BTFS. Seismic slip vectors suggest that the east BTFS may represent a mature transform fault system accommodating the plate motion. At the same time, the western BTFS is immature, as its reorganization is still ongoing (Ren et al., 2023).

Previous marine observations include regional hydroacoustic observation (Fox and Dziak, 1999) and a one-year-long OBS experiment between 24 September 2012 and 2 October 2013 (Blanco Transform OBS Experiment, network code X9, (Kuna et al., 2019; Nábělek and Braunmiller, 2012). Regional studies of seismicity and focal mechanisms constrained from onshore seismic networks (Braunmiller and Nábělek, 2008) revealed that the BTFS is segmented, with different sections exhibiting varying levels of seismic activity. These segments are often separated by step-overs or bends in the fault, which can act as barriers to earthquake rupture propagation (Fox & Dziak, 1999). To the East, the Gorda Depression with pull-apart basins, sub-parallel faults and seismicity down to ~9 km is found. The adjacent segment is characterized by a ~150 km long and straight transform segment with deeper faulting down to

~12 km depth, while most of the seismicity occurs at depths shallower than 10 km and above the 600° isotherm consistent with the brittle-ductile transition (Dziak et al., 1991; Roland et al., 2012; Ren et al., 2023).

## 3. Data

The BTFS was instrumented with a dense network of 55 ocean-bottom-seismometer (OBS) stations between September 2012 and October 2013 (network code X9, Kuna, 2020; Nábělek and Braunmiller, 2012) and an average station spacing of ~25 km. Thirty-one stations had broadband Güralp CMG-3T seismometers, and the other 24 stations had short-period Mark L-28LB seismometers with differential pressure gauges (Scripps Institution of Oceanography DPG). Besides station BB270, 54 OBS record continuously four channels.

## 4. Methods

### 4.1. Phase Picking using different picking algorithms

We determine P- and S-phase arrivals using four machine-learning (ML) pickers trained to create a high-resolution local seismicity catalog. We picked the complete dataset with the two picking algorithms trained with 3-component land data (EQTransformer from Mousavi et al., 2020), and Phasenet from (Zhu and Beroza, 2019). Two other pickers were trained on OBS data (BLUEQTransformer and BLUEPhasenet from Bornstein et al., 2024). The OBS picker uses the hydrophone channel as an additional 4th component. They are based on transfer learning of the land data trained pickers and an OBS training dataset (Bornstein et al., 2024), which included a small number of manually picked onsets (3,811) from the BLANCO OBS dataset. The difference between the picker and human picks is shown in Figure S1.

We only used picks with confidence thresholds exceeding 0.6 for all picking algorithms to avoid outliers and optimize the hit rate. Compared to the events picked by human pickers on the OBS data, this resulted in less than 3% mispicks for P phases and a miss rate of ~20% (Bornstein et al., 2024). There are less than 1% wrong picks for S phases but a miss rate of >50%. Based on the events database, the OBS pickers perform similarly to human experts and outperform phase pickers based on land data only. However, there is limited experience applying the ML OBS picker to continuous data (Bornstein et al., 2024).

### 4.2 Event Association

The picks from the three different pickers (EQTransformer (Mousavi et al., 2020), Phasenet (Zhu and Beroza, 2019), PickBlueEQtransformer, PickBluePhasenet (Bornstein et al., 2024) were associated to events using REAL (Zhang et al., 2019). The association was done with the minimum 1D velocity model for the BTFS (Ren et al., 2023). We found that associating with a constant vp and vs model or the minimum 1D velocity for the BTFS worked equally well.

4.3 Absolute Locations

We employ the non-linear oct-tree search algorithm (NonLinLoc, Lomax et al., 2000) to determine absolute hypocenters. Travel times are computed using the finite-difference solution to the Eikonal equation (Podvin and Lecomte, 1991). This oct-tree algorithm offers more reliable location uncertainty information than linearized inversions by exploring the probability density functions (PDF) of individual events. The maximum likelihood location is selected as the preferred location. NonLinLoc estimates a 3D error ellipsoid from the PDF scatter samples for each event. The ellipsoid with the uncertainties (68% confidence) was projected onto the latitudinal, longitudinal and vertical directions (Lange et al., 2012) to generate uncertainties for the geographical system. Station corrections for P and S phases, accounting for localized deviations from the a priori model, were derived from the average residuals at each station (Husen et al., 1999). This was done by repeatedly locating the events with NonLinLoc. After each location with NonLinLoc, we updated the station corrections from the last iteration. We found that after 4 iterations, the mean of the station corrections converged to a minimum (Figure S2). Station corrections were estimated using a high-quality subset of 1,380 events (GAP<=200°, more than 14 onsets and RMS smaller than 1 s. We classified events into five quality categories (class A-D, A being the best) using a slightly modified scheme of (Husen and Smith, 2004).

We selected for each of the four catalog events with more than 8 stations and an RMS smaller than 1 s. Comparing the three catalogs showed that the events located differ between the catalogs, in particular for small events. The benefit of having different pickers is that this can be used as a proxy for onset uncertainties by comparing the different picks for the same onset. Therefore, we combined the different catalogs to create a final dataset. We created quality groups 0, 1, 2, and 3 for the picks in case more than one pick was available. If BLUEQTransformer, BLUEPhasenet, or EQTransformer picked a phase, we estimated the quality from the time difference between the picks and used the onset time from the PickBlueEQTransformer picker. We use the standard quality label (weight) according to the time difference between the pickers (0: 0.05 s; 1: 0.1 s; 2: 0.2 s; 3: 0.4 s). In case there was only one onset, we used the confidence value of the picker as a proxy for pick weight (0: >0.9; 1>0.8>;2>0.7;3>0.6). We find that using weights focused the hypocenter locations better for the final catalog (Movie S1 and S2) and reduced the RMS (Table T1 in T1). Furthermore, using weights decreased the catalog's RMS by approximately 11%, suggesting a better data fit.

Furthermore, the final catalog based on the combination of all available picks significantly improves the number of events in the dataset, which has ~14,700 events compared to the ~8.000 events from the catalogs based on phase picker only. Further, we deleted five events far outside the working area and 625 events (5%) with RMS larger than 1 s. The remaining dataset has 14,139 events. By carefully inspecting, we found that 92 events (0.65%) occurred within 10 seconds of each other. The reason is that they were combined in two separate events during the merge of the three catalogs, which was based on the origin time of the catalogs. Slightly different origin times after the association wrongly resulted in two events during the association of the two catalogs. After carefully inspecting the 92 events close in time, we kept the event with the larger number of station observations and where the closest station had a smaller distance to the epicenter.

Lastly, we looked at teleseismic events that might have been wrongly associated and located in the network. Since a local event should always be picked first at the closest stations and teleseismic events should be picked across the whole network, we counted how many stations had less than three picks within the next 50 km epicentral distance. Only 32 (0.2%) events did not fulfill this classification. Manual inspection showed that these were still local events, so we did not use this criterion for the final catalog selection in contrast to a previous study for the BTFS (Kuna, 2020). EQTransformer was reported to miss strong earthquakes in OBS data (Gong et al., 2022). Using the OBS Pickers and the BTFS, we found that our processing scheme detected most of the events reported by NEIC (Figure S3). The final absolute catalog has 13,503 events based on 243,667 arrival times (139,361 P phases, 104,306 S phases) with a mean uncertainty of 500 m for the horizontal coordinates and 1 km in depth (Figure S4).

### 4.4 Relative Locations

We use the relative location program HYPODD (Waldhauser and Ellsworth, 2000) to obtain accurate relative earthquake locations. Waveform cross-correlations are calculated on the vertical component filtered between 2 and 10 Hz using 2.5 and 3.5 s long windows for the P- and S-waves, respectively. The waveforms under consideration start 0.5 s before the picked onsets. We use correlation coefficients larger than 0.7 and the square of the correlation coefficient as a weight (Chalumeau et al., 2024). HYPODD seems unable to correctly locate all the events in the working area (likely due to the number of events exceeding 10,000 events in one cluster), so we separately relocated the events east and west of 128,9° longitude. Due to the high number of stations and events, 90% of the absolute seismic catalog was part of a cluster, resulting in a relative catalog with 12,731 epicenters. The uncertainties of the relative catalog are likely smaller than 500 m since the events off-axis cluster all fall within such a small radius (Figure S5). A comparison of the relative with the absolute locations is shown in Figure S6, and the temporal distribution of the absolute and relative catalogs are shown in Figures S7, S8, S9 and S10. The seismicity in time is shown in the Movie S3.

### 4.5 Magnitudes

Moment magnitudes (Mw) and local magnitudes (Ml) were estimated using the automated procedure of (Ottemoller and Havskov, 2003) integrated into the SEISAN software (Havskov and Ottemoller, 1999). For the local magnitudes (Ml), we used the standard formula and values for geometric spreading and attenuation (Hutton and Boore, 1987). The magnitudes from the absolute catalog are based on 257,850 spectral and 101,386 amplitude estimates. The resulting catalog lists events with Mw between 1.2 and 5.3 (Ml between 1.3 and 5.0) (Figure S11). Using the software package ZMAP (Wiemer, 2001), we estimated a magnitude of completeness (Mc) of 2.1 and a b-value of 1.30.

### 4.6 Repeaters

To determine repeaters (Nadeau and Johnson, 1998; Uchida and Bürgmann, 2019), we cross-correlated 0.5 s seconds before until 15 seconds after the P arrival. In general, we use the widely used conversion (Uchida and Bürgmann, 2019) from moment magnitude to slip using the formula from (Nadeau and Johnson, 1998). First, we cross-correlated all stations with phase picks with all events with an epicentral inter-event distance of 4 km and within an epicentral distance of less than 50 km. Before cross-correlation, the seismograms were detrended and filtered between 1 and 20 Hz. For the repeater study, we used the whole relative catalogue. To avoid similar events (Gao et al., 2021), we used more rigorous requirements for grouping event pairs into repeater groups. For grouping of repeaters, we only considered event pairs with cross-correlation coefficients larger than 0.95 found for at least two stations. The requirement that each event pair needs at least two high CCs makes the group selection less dependent on the station azimuth and epicentral distance.

### 4.7 Local Earthquake Tomography

We invert two-dimensional (2D) velocity models along the transform fault system using local earthquake tomography (LET) using the widely used inversion code SIMUL2000 (Thurber and Eberhart-Phillips, 1999; Thurber, 1983). In the damped least-squares inversion, the vp velocities and the ratio between P and S wave velocities (vp/vs) are inverted from the observed travel times of P and S arrivals. The velocity model is represented by velocity values specified on a rectangular grid of irregularly spaced nodes. Following common practice, we applied a staggered inversion scheme inverting a two-dimensional (2D) velocity model vp using a minimum 1D velocity model (Kissling et al., 1994) from (Ren et al., 2023) from a brute force search of different one-dimensional input models as start model, followed by the inversion of the 2D vp model. For the inversion of the vp/vs model, we fixed (i.e., highly damped) the vp model and used a constant vp/vs ratio of 1.77 derived from Wadati diagrams (Figure S10) as the starting model. All events were relocated before each iteration, and the damping values were 200 and 500 for the vp and vp/vs inversion, respectively. More details on the steps involved in the inversion steps are described in (Lange et al., 2018). For each inversion step, the RMS decreased. For the last inversion (vp/vs), the overall RMS decreased by 23% to 0.281 s.

**5 Results**

### 5.1 Seismicity catalogue

The final relative and absolute catalogs have 13,503 and 12,731 local events (Figures 1, 2, and 3), respectively, and a magnitude of completeness (Mc) of 2.1. The seismicity reveals detailed information on the deformation along the BTFS, including smaller step-overs, transtensional structures and almost aseismic segments.

Three primary characteristics of the catalogs are as follows:
1.) Increasing hypocentral depths are observed within the first 50 km from the ridge axis, (Figure 2c).
2.) In the western section of the BTFS (Figure 2c, between 50 km and 170 km profile distance), seismic events are predominantly shallow, occurring mostly at depths less than 10 km below the seafloor (e.g., 13 km below the sea surface) in the oceanic crust. Seismicity for the western BTFS is found to occur at deeper levels, with most of the seismicity up to 16 km below the seafloor and related to the oceanic crust and mantle.

3.) The western BTFS exhibits a highly segmented seismicity pattern, characterized by more events (Figure 2C) and a lower b-value (Figure S11) than its eastern counterpart.

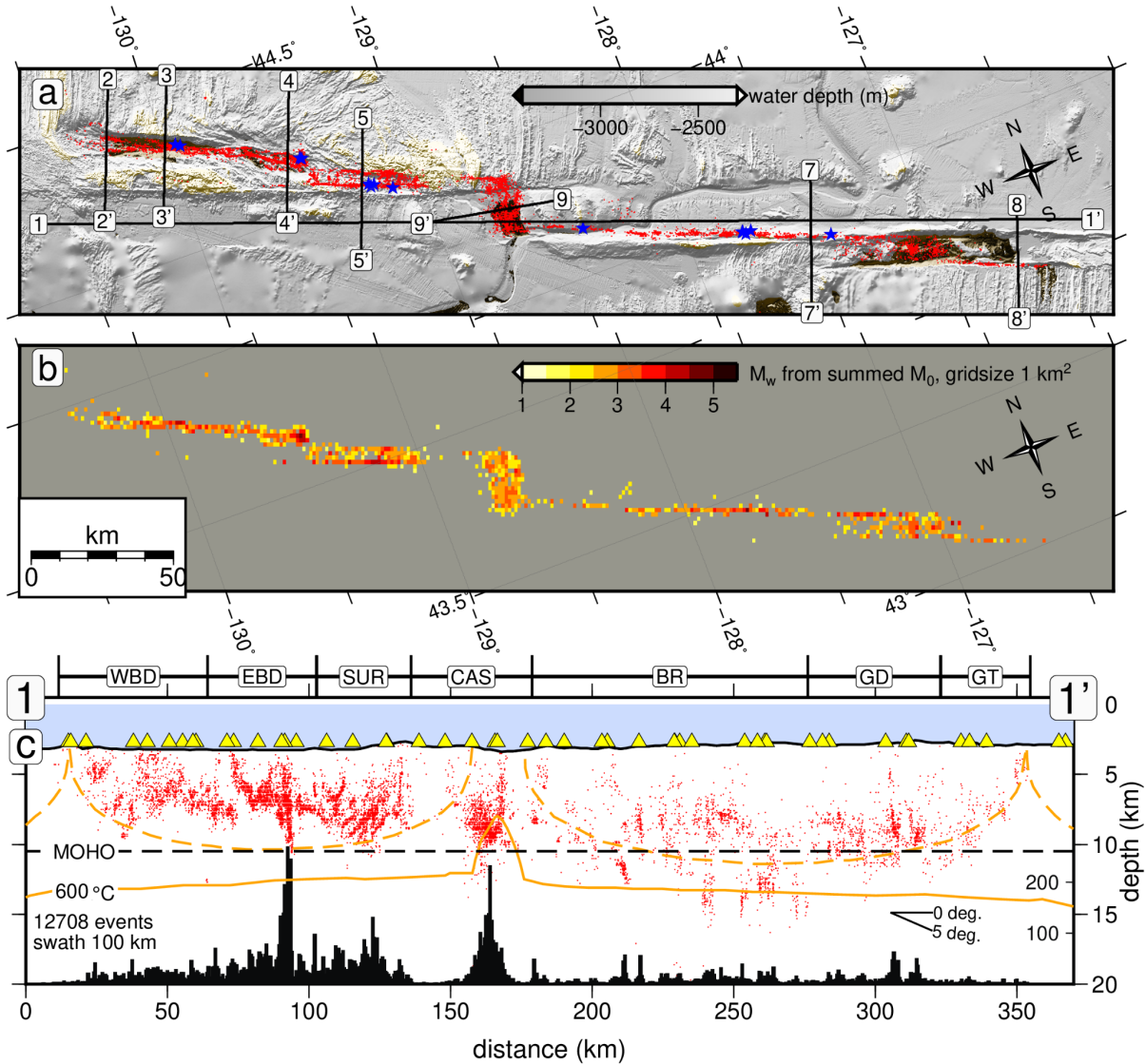

**Figure 2:** Microseismicity of the relative earthquake catalog in map view (a), as magnitude moment densities (b) and as vertical profile along the BTFS (c). (a) Events larger than Mw 4.5 are shown with blue stars. Bathymetry from Ryan et al., 2009. (b) Colour squares indicate the cumulative moment release estimated from the local seismicity, converted to equivalent moment magnitude for tiles of 1 km² size. For the summation of the seismic moments, we assumed the moment to be located at the epicenter. (c) The histograms show the number of the events along the sections together, and the number of events is labeled to the right inside panel c. The Dashed purple line shows the Moho discontinuity inferred from a minimum 1D velocity model (Ren et al., 2023). The dashed orange line indicates the 600°C isotherms from a half-space cooling model, and the orange line indicates the 600°C isotherms considering hydrothermal cooling (Roland et al., 2010).

manuscript submitted to *Journal of Geophysical Research: Solid Earth*

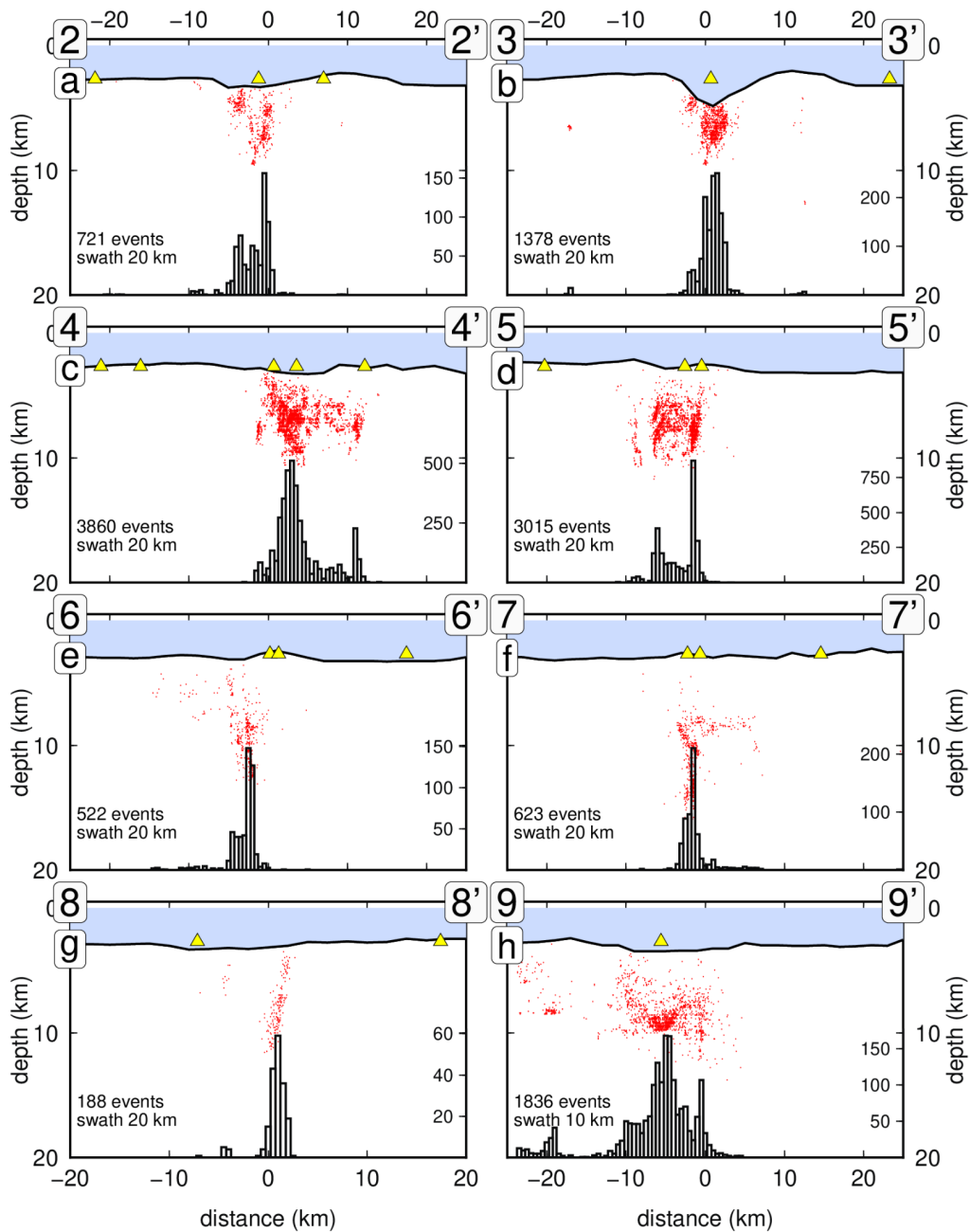

**Figure 3:** Cross-sections of the seismicity. The locations of the vertical profiles are shown in Figure 2. Events are shown within 20 km (10 km) of either side for profiles.

In the following, the seismicity distribution is described for the geological units (Figure 1, Embley and Wilson, 1992) from West to East:

– West Blanco Depression (WBD): The westernmost 120 km of the BTFS are characterized by an elongated basin, the West Blanco Depression (Figure 4a). Within this region, seismicity predominantly occurs in a zone approximately 10 km wide (Figure 3a & b). As seismicity approaches the intersection with the Juan de Fuca Ridge, it appears to bifurcate into two parallel clusters that align with the direction of the ridge (Figure 4a). Despite this curvature towards the Juan de Fuca Ridge, seismic activity becomes notably sparse along the ridge. The OBS should have been able to detect seismic activity related to the Juan De Fuca ridge since the ridge-transform intersection is still covered by the network, indicating a significant difference in moment release from microseismicity between the transform and the adjacent spreading segments.

– Parks Plateau (PP): Located approximately 20 km south of the WBD, the Parks Plateau is associated with only individual seismic events. These events appear to align along the bathymetric ridge, and their sparse distribution can be traced on the map view until reaching the Surveyor Depression (SUR) farther to the east (Figure 4a).

– East Blanco Depression (EBD): Seismicity occurs in at least two sub-parallel clusters within the EBD. The fine structure of the clusters partly shows step-overs indicating a complex deformation.

– Surveyor Depression (SUR): Significant seismicity is observed along the SUR, characterized by two parallel, closely spaced clusters separated by approximately 7 km (Figure 3d). Notably, the southern strand extends further westward, continuing towards the EBD. The area between the SUR and the Cascadia Depression exhibits nearly aseismic behavior (Figure 2c).

– Cascadia Depression (CAS): The CAS is a trough with a water depth of ~3500 m. West of the CAS, an ~12 km long zone with sparse seismicity, almost aseismic, is observed (Figure 4b). The CAS itself is the most seismically active region of the BFTZ with ~1,800 events across a profile of 20 km width, peaking in the center of the depression (Figure 4h). Repeaters are located mainly within the basin. The seismicity patterns suggest north-south trending faults (Figure 4b), likely indicative of extension related to spreading or initiation of spreading due to the step-over of the BTFS.

– Blanco Ridge (BR): Seismicity exhibits reduced complexity along the linear BR. Events do not align directly with the BR morphology but appear to shift clockwise by approximately 1° with seismic activity to the east occurring north of the BR (Figure 4b, at 128.5°). East of 128.25°W longitude a subtle directional shift of seismic activity away from the BR and towards the north is observed resulting in an ~5 km northward displacement of seismicity close to the CAS (Figure 1 and 4b). We attribute this seismic-bathymetry misalignment to a reorganization of the stress field. Seismicity reaches up to ~14 km below the seafloor, consistent with previous studies (Kuna et al., 2019; Kuna, 2020), but displays less clustering, likely due to smaller hypocentral uncertainties of our catalog. Along the more simple and linear eastern transform fault segment of the BR, the deformation is within +/-1.5 km of the fault trace and, hence, very focused (Figure 3f and 3g). Taking into account the horizontal uncertainty (68% confidence) of 600 m (<500 m) for the absolute (relative) catalog (Figures S5 & S6), this indicates a minimal fault zone width of less than 1 km.

– Gorda Depression (GD): Seismicity within the GD is predominantly focused in the central, deepest part of the depression, related to two seismicity clusters (Figure 2c, between profile distances 320 km and 335 km). The GD is a pull-apart basin (Braunmiller and Nábělek, 2008). The ~20 km step-over of the overall faulting results in transtensional deformation of the same width (Figure 4C). Although the morphology of the BR continues west of the GD, seismicity in this area is sparse, with only a few recorded events.

– Gorda Transform (GT): East of the GD, seismicity is linear but displaced southward with the majority of events concentrated along the southern limit of the GD (Figure 4c). Similar to the BR, the seismicity shows a focused linear distribution, with a damage zone around the GT measuring approximately +/-1.5 km wide for the main fault. Sparse seismicity is located 5 km north of the GT (Figure 3g, at +5 km distance).

– Gorda Ridge (GR): No seismicity was detected along the GR despite the network coverage at the transform-ridge corner. Seismic activity abruptly ceases at the ridge-transform intersection.

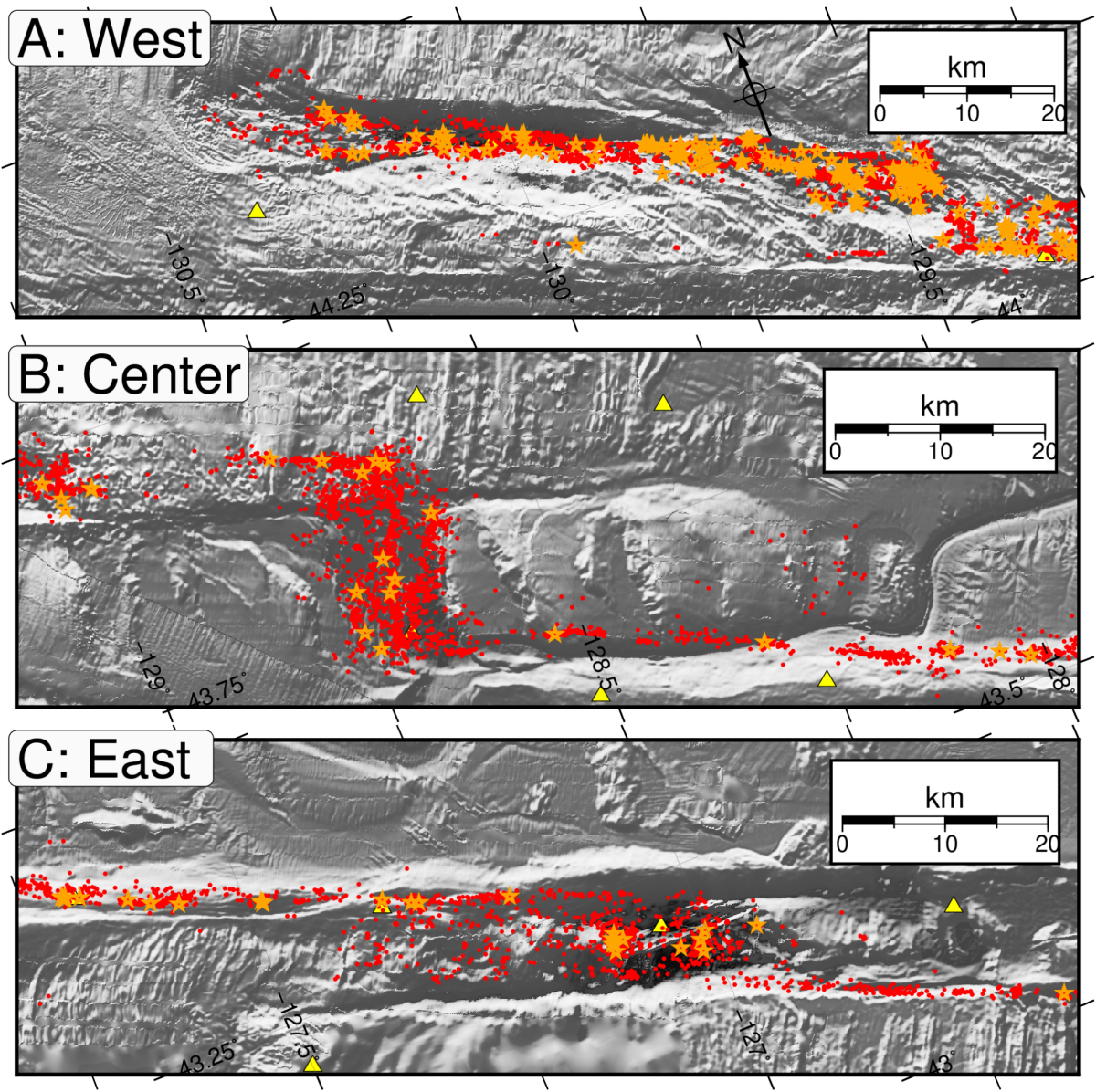

**Figure 4:** Closeups (map views) from the relative seismicity catalog of the Western (a), Central (b) and Eastern (c) part of the BTFS. Seismicity is shown with red dots, repeaters with orange stars. All three panels have the same scale. The location of the closeup views is shown in Figure 1. OBS stations are indicated with yellow triangles.

5.4 Repeaters

We found 278 groups of repeaters, including 211 doublets and one repeater with 7 event members (Figure 5a). The slip from repeaters ranges between 11 and 61 cm with a mean of 21 cm and we cannot identify clustering in time (Figure 6). There are significant variations in the spatial distribution of repeaters along the fault (Figure 7). At first glance, the spatial distribution of repeaters along the BTFS closely resembles that of the seismicity catalog. Similar to the seismicity, the depths of the repeaters increase from west to east, mostly confined to a depth interval shallower than 12 km (Figure 8a). For identifying repeaters, the interevent separation needs to be smaller than the source dimension of the larger event (Gao et al., 2021). Therefore, we checked all repeater groups and found a mean difference of 1.7 km and 1.1 km to the cluster centroids for the horizontal coordinates and depths, respectively.

The largest repeater group, with 7 event members within 6 days at the beginning of March 2013, had a total slip of 61 cm (Figure 6). Figure S12 shows the seismograms of this repeater group. Overall, most repeaters have inter-event times of less than 30 minutes (Figure 5a). With longer inter-event times, the number of event pairs declines significantly. Notably, no inter-event times exceed ~165 days despite the deployment time of one year. In a first-order order, the number of events along the BTFS is similar to the number of seismicity (Figure 6a).

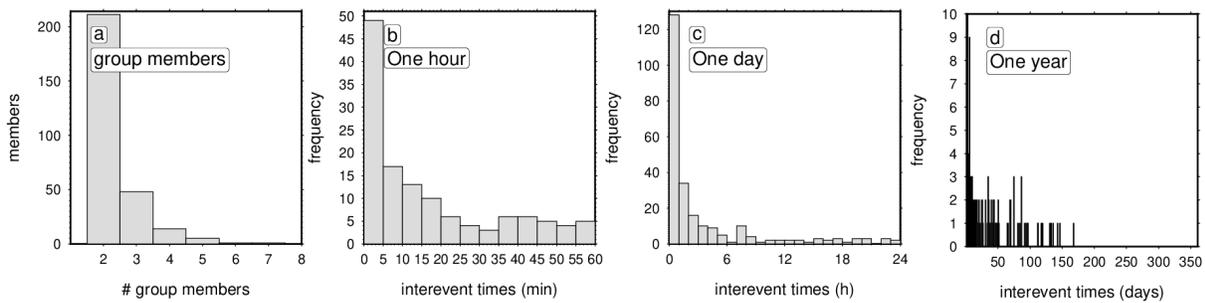

**Figure 5:** Histograms of repeater group events and inter-event times of the repeaters. (a) Histogram of the number of events per repeater. (b) Histograms for the interevent times of one hour. (c) Histogram for one day. (d) Histogram for one year. To visualize the decrease of repeaters in time, the y-axis in this panel is limited to 10 repeaters.

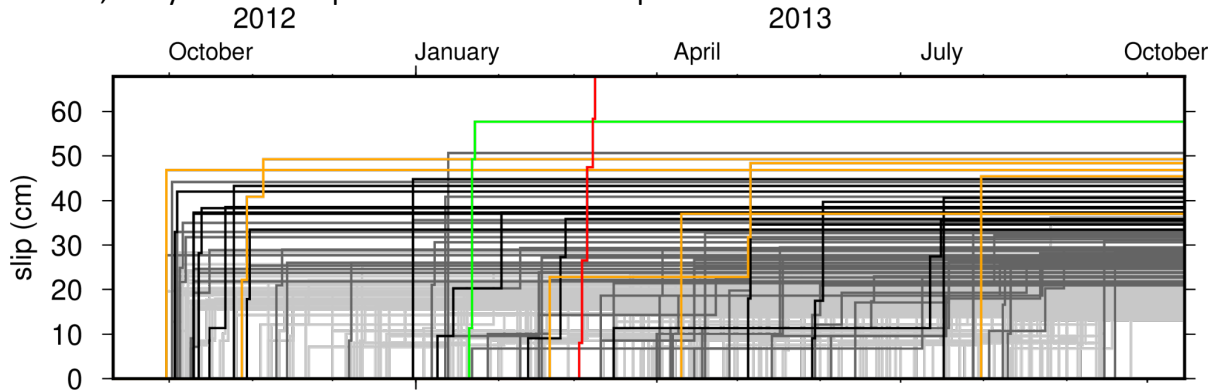

**Figure 6:** Slip from repeaters in time. The colors indicate the number of repeater members in each group.

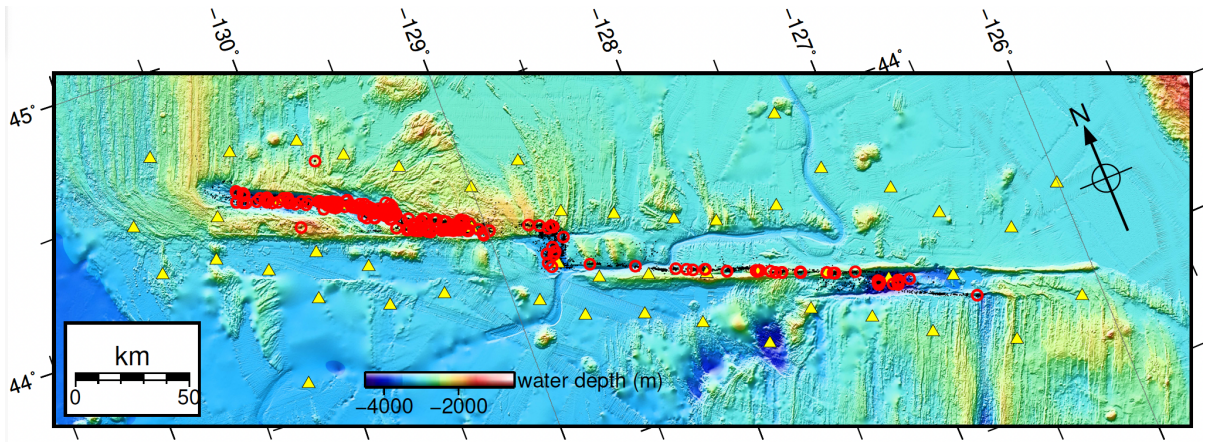

**Figure 7:** Map showing the distribution of repeaters (red circles). Relative epicenters of the relative catalogue are indicated with black dots.

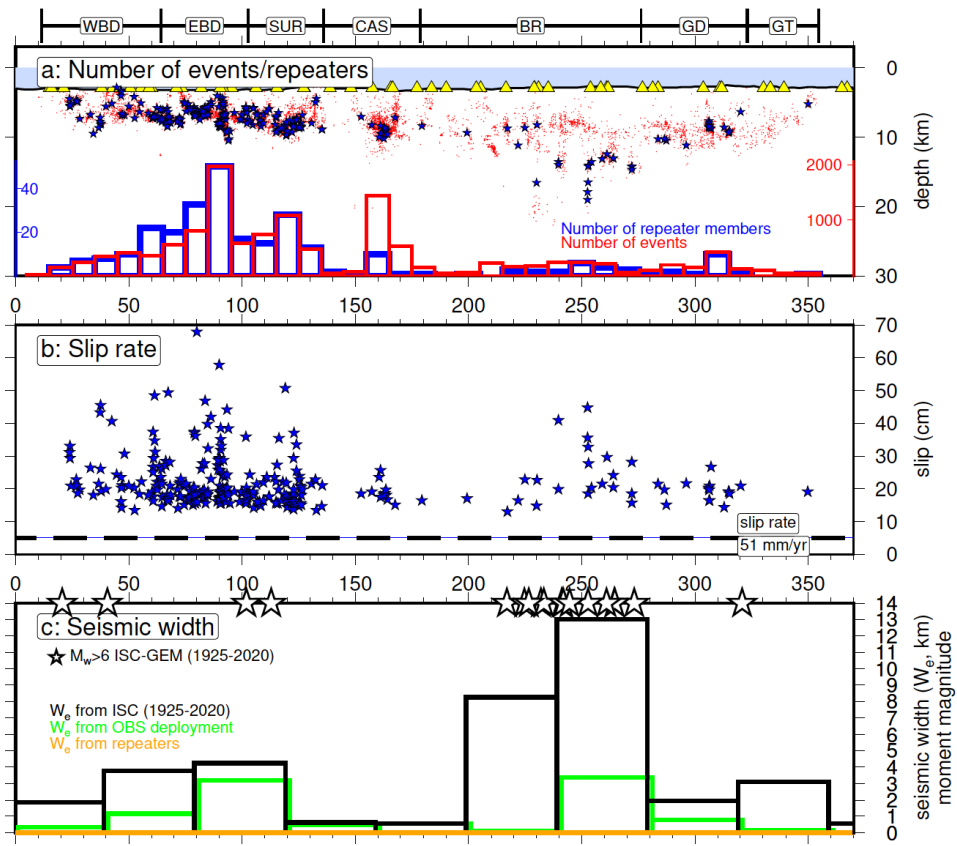

**Figure 8:** Distribution of repeaters along the BTFS. (a) Histogram of the number of repeaters (blue) and histogram of the number of relative seismicity catalogue along the BTFS (red). Blue stars indicate repeaters, red dots indicate seismicity from the relative catalogue. Geologic units are labeled on the upper side. (b) Slip of repeaters shown with blue stars. The slip rate of 51 mm/yr (DeMets et al., 2010; Wolfson-Schwehr and Boettcher, 2019) is shown with the dashed line. (c) Effective seismic width and seismicity (Boettcher and Jordan, 2004) for the BTFS for the local deployment (green), the ISC catalog (1925-2020, black) and the repeating events (orange).

5.5 Local Earthquake Tomography

The final two-dimensional vp and vp/vs models are shown in Figure 9. For the uppermost 3 km of oceanic crust, the vp velocities are around 4 km/s with higher velocities (~5.2 km/s) related to the West Blanco Depression (WBD) and the CAS. The lower crust has four regions of elevated vp velocities, as can be seen from the curved 6 km/s contour line in Figure 9. Please note that the velocity increase at ~10 km depth below sea level stems from the vp velocity contrast introduced by the minimum 1D velocity model (Ren et al., 2023). The vp model suggests little lateral changes of the MOHO to the 1D velocity model, likely related to little lateral MOHO depth changes along the strike. The vp/vs model shows mostly elevated vp/vs ratios exceeding 2, particularly for the aseismic zone East of the CAS and at 127° longitude.

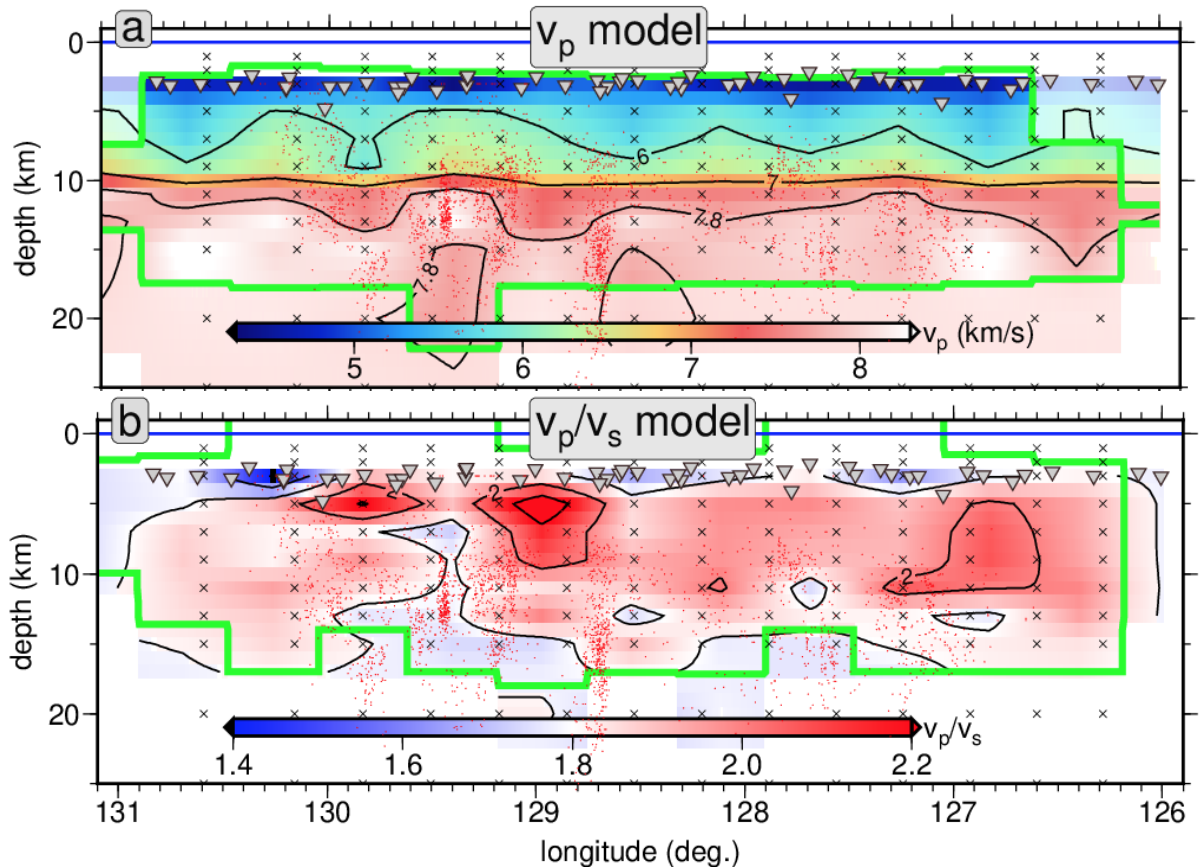

**Figure 9:** 2-D tomographic velocity models along strike of the BTFS for vp (a) and vp/vs (b) velocity models. Regions with good resolution are encircled with a green line. Red points hypocenters (located in the 2D vp/vs velocity model) and grid nodes are shown with crosses. Stations are indicated with inverted triangles. The black lines indicate contour lines of the velocity model and the blue line indicates the sea surface. Please note that the velocity increase in 10 km depth for the vp velocity model might be given as a-priori information to the inversion since this mimics the velocity increase of the MOHO from the minimum 1D velocity input model.

## 6 Discussion

Modern phase-picking algorithms based on machine learning applied to a dense network of 54 OBS along the BTFS result in ~1,3000 events, revealing the current deformation of a transform fault system in high-resolution detail. The catalog shows various linear features revealing the seismic deformation of the BTRS with high resolution (Figures 2-4).

### 6.1. Deformation from seismicity

The rapid development of phase pickers using machine learning facilitates the picking of phases of a large number of stations and events. These pickers were retrained for OBS data to use four channels, including the hydrophone channel (Bornstein et al., 2024). To create the catalog, we use phased pickers optimized on land data (Mousavi et al., 2020; Zhu and Beroza, 2019). Since there is limited experience in applying automated phase pickers on OBS data, we investigated the differences between the catalogs, including combining phase pickers for a unified catalog. Combining the picks of different pickers, including arrival time uncertainties based on time differences derived from the different pickers and the use of station corrections (Figure S2), further improved the catalog. Upgrading picks with similar pick times of different phase pickers results in more focused seismicity. The picker trained with four components of OBS data (PickBlue, Bornstein et al., 2024) performs better than the pickers trained on land data. Concerning the number of events, we find 30% more events than traditional processing using STA/LTA filters for detection (Kuna, 2020).

Seismicity is mostly focused along faults outlined by seafloor morphology and the hypocenters cluster within 2.5 km of the primary fault trace. This is in contrast to the Discovery Transform (Wolfson-Schwehr et al., 2014) where the seismicity is clustered within 5 km of the primary fault trace or shows events some km away from the main fault (Gong et al., 2022). We relate the difference in the hypocentral uncertainty of our catalog. Taking into account the uncertainties of the hypocentral locations, seismicity reveals similar widths of the damage zone around the fault as for continental strike-slip faults such as the San Andreas fault (Ross et al., 2017; Shearer, 2002; Waldhauser and Ellsworth, 2000) or to normal faults (Valoroso et al., 2014). 50 km away from the Ridge-Transform intersections, seismicity fades out in terms of frequency and faulting depth, likely related to the increasing temperatures towards the ridges (Braunmiller and Nábělek, 2008; Roland et al., 2010; Wolfson-Schwehr and Boettcher, 2019).

We find sparse seismicity towards the Gorda Ridge and the Juan de Fuca Ridge (JdFR). For the JdFR, this would be in line with the observations using acoustic direct path ranging between 1994 and 1996 (Chadwell et al., 1999) who could not resolve spreading in the center of the spreading system at the boundary of the OBS network (latitude 44°40'). The JdFR is characterized by a very low seismicity rate (Nedimović et al., 2009). We find one off-axis event cluster (44,4°N/129.68°W, Figure 2 & S5), likely the same cluster that was observed using hydroacoustics in 2008 with 600 events and was suggested to be related to internal plate deformation related to the propagator wakes in the region (Nedimović et al., 2009). The cluster was active for only 84 hours (2013-06-27 13:34:52.90 until 2013-06-30 21:05:20.20 UTC) with magnitudes 1.4-2.4 Ml. The three closest stations with distances of 10,13,24 km registered P-S onset times of 1.9, 2.1, and 3.2 s. Most seismicity occurs in a ~1 km2 large region (Figure S5), suggesting hypocentral uncertainties of less than 500 m for the relative catalog.

6.2. In-situ slip from repeater

Repeating earthquakes (repeaters) that rupture the same fault area (patch) are interpreted to be caused by repeated accumulation and release of stress on the seismic patch in a creeping area (after Uchida and Bürgmann, 2019). Therefore, repeaters allow the estimation of fault slip rates in-situ at depth from their recurrence (Nadeau and McEvilly, 1999), and the repeater represents the full in-situ slip of its aseismic surrounding (e.g., Uchida, 2019). It is long known that 85% of the plate motion for transform faults is released aseismically (Brune, 1968, Boettcher and Jordan, 2004). The long-term slip rate for the BTFS from the NUVEL model (from magnetic lineaments) is 51 mm/yr (DeMets et al., 2010; Wolfson-Schwehr and Boettcher, 2019). During the OBS deployment, the mean seismic width (Boettcher and Jordan, 2004) is 1.1 km, resulting in ~20% aseismic slip for most of the BTFS (assuming faulting depth down to the Moho, Figure 8).

Most repeaters are related to faulting processes in the oceanic crust. However, 13 repeaters beneath the Blanco Ridge (at 260 km profile, Figure 8a) occur significantly below the MOHO and at depths up to 20 km indicating faulting in the oceanic mantle. This region is also characterized by stronger events and the largest moment release (Figure 8c). Braunmiller and Nábělek (2008) found a similar behavior using slip rate estimates from global catalogs and suggested that this section is fully locked. We agree with this finding, particularly as this region has hosted 12 earthquakes larger than magnitude 6 since 1925 and its seismic width (Boettcher and Jordan, 2004) exceeds with 13 km the crustal thickness (Figure 8c). This large seismic width indicates significant moment release and therefore strain buildup related to a locked fault. We are aware that the uncertainties of the historical catalogue include uncertainties for depth and magnitudes. However, due to possible missing events in the catalogue, the seismic width from the historical catalogue would be underestimated.

We note that the b-value is slightly larger for the Western part of the BTFS than the East BTFS (Figure S11). The small events for the West do not contribute significantly to the seismic moment release (Figure 8c) as suggested previously (Braunmiller and Nábělek, 2008).

The mean seismic width (Boettcher and Jordan, 2004) of the BTFS during the deployment (Figure 8c) is 1.1 km and 3.7 km for the long term seismicity catalog of 95 years from the ISC-GEM catalog (Di Giacomo et al., 2018; Storchak et al., 2015, 2013). The differences of seismic widths are mostly related to the Blanco Ridge. The western BTFS seismic widths are similar for the OBS deployment and the historical record (1-3 km), suggesting a substantial creep of ~60%.

The slip rate from repeaters ranges between 11 and 61 cm/yr (Figure 6 and 8b) clearly exceeding the geologic strike-slip rate of 51 mm/yr. Following the widely accepted suggestion that the repeater reflects the full slip of the fault, this indicates that each repeater group releases the slip accumulated three or more years from the BTFS fault slip rate. Therefore, we interpret the fault behavior as many small locked patches within the overall mostly aseismic fault system. Kinematically, the fault appears aseismic and accumulates stress at the locations of the repeaters for many years before releasing their accumulated stresses, indicating a stick-slip movement with repeat times of years. Since the repeaters appear not to be related in time but rather occur randomly during the deployment (Figure 6), we suggest that the fault consists of many small patches which have at least the size of the repeater rupture areas of some hundred meters (corresponding to a Mw 2-3 earthquake). Between the moment release from repeaters, most seismic moment is released by creep processes.

Overall we notice an absence of crustal repeaters related to the locked patch beneath the BR (Figure 8a, between 220 and 270 km distance), but repeaters occur exclusively at mantle depths exceeding 14 km. Based solely on seismicity, Kuna et al. (2019) suggested as well that this segment is coupled, while the fault in the mantle, shallower than the depth of the 600°C isotherm, creeps episodically and produces seismic swarms and foreshocks. The repeaters distribution is in line with Kuna et al.'s interpretation, which is that for the Blanco Ridge the crustal part is loaded from creep occurring in the mantle, while large earthquakes regularly break the crust and the mantle. The distribution of repeaters, together with the seismic width, indicates a partly locked (~30%) BTFS related to the WBD, EBD and SUR and the GD and GT towards the East (Figure 8). The Eastern Blanco Ridge appears to be 95% creeping, while the Western Blanco Ridge is fully coupled.

### 6.3. Velocity Structure and Serpentinisation

The tomography clearly shows elevated vp/vs values for most of the BTFS (Figure 9). The regions with reduced (e.g. vp/vs<1.78) ratios are related to bathymetric depths, for example, towards the East associated with the uppermost ~3 km crust of the West Blanco Depression and the Gorda Depression. The middle of the Blanco ridge (~128.5°) also shows lower vp/vs ratios of 1.7. Figure 10 shows a modified Wadati plot using the dense seismicity beneath the stations allowing to map vp/vs (in low resolution) along the BTFS. Although this modified Wadati plots result in vp/vs is not as accurate as from the inversion, this figure is closer to the raw data since it only makes use of the picked onsets and origin times (Haberland et al., 2009), allowing to retrieve information on vp/vs ratios directl from the onset times. Overall, our vp/vs ratios are significantly higher than those vp/vs values found for the moderate ratios along the Gofar transform (Liu et al., 2023; Guo et al., 2018). The vp/vs ratios for the Romanche transform are below 1.85, and the ratios are very low at the ridge transform intersection (Zhiteng et al., 2023).

High vp/vs ratios in the crust are indicative of fluid filled cracks (Wang et al., 2012; Popp and Kern, 1994), which is in agreement with seismic activity at crustal depth. For the mantle, however, serpentinized material is characterized by clearly elevated vp/vs and reduced vp velocities (Carlson and Miller, 2003). In particular, serpentinization favors aseismic sliding (Reinen et al., 1991), and the BTFS was hypothesized to be serpentinized (Kuna et al., 2019). For the region with the most elevated vp/vs ratios (east of the CAS, at ~129°), a vp/vs ratio of 2.1, together with a vp velocity of 6.2 km/s, might indicate a serpentinite content of maximal 60% (e.g. 7% volume content water, Carlson and Miller, 2003, their figure 2). On average, we roughly estimate 30% serpentine volume content (e.g., 3% water, based on vp/vs=2, vp=6 km/s). The elevated vp/vs ratios are promoted from highly fractured rocks, which can further increase the vp/vs and are suggested to further promote serpentinization from seawater entering the crust through these pathways (e.g., Lefeldt et al., 2009; Wang et al., 2022). Serpentine is velocity weakening at fast loading velocities and velocity strengthening under slow loading conditions, allowing for the propagation of slip into serpentinized patches (Reinen et al., 1991). If true, at least the region around each repeater could be seen as a velocity-weakening patch breaking after some years and converting into a velocity-weakening patch with slip release.

The serpentinization of the BTFS explains why no earthquakes larger than Mw 6.4 were observed since the serpentinization and fluids promote seismic and aseismic, limiting the accumulation of stress.

The velocity models do not show the different seismicity behavior of the Western and Eastern BTFS, indicating that the higher faulting depths to the West are not directly linked to velocity structure and might, therefore, be controlled by the larger depth of the fault in the West compared to the eastern BTFS. Since the Eastern transform faults are more mature and older than the Western segments (Ren et al., 2023), this might have also resulted in larger faulting depths.

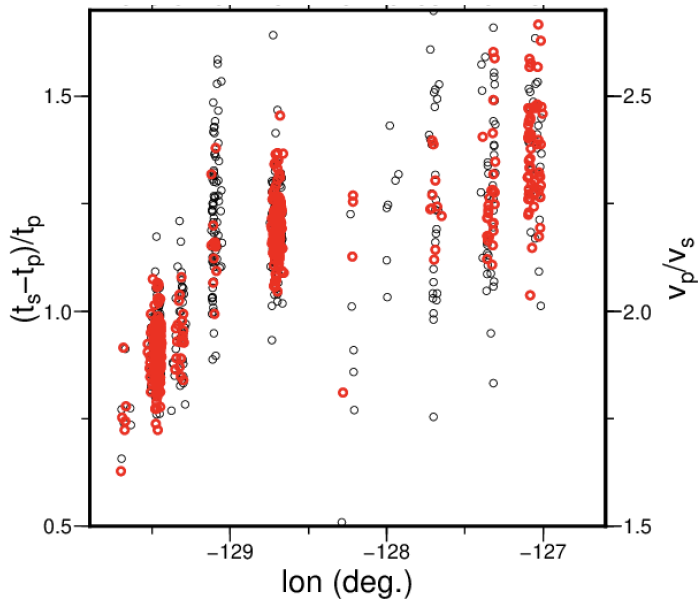

**Figure 10**: The S-P traveltimes (normalized to P wave traveltime) plotted against event longitudes for all observations with less than 4 km epicentral distance from the OBS and event depth of more than 5 km (e.g. sampling sub-vertical rays). Events with P and S arrival with the highest quality are shown in red. $(t_s - t_p)/t_p$ values for the OBS stations west of 129° source longitudes indicate an average lower $v_p/v_s$ towards the East. The station names are labeled on the upper side.

**7 Conclusions**

We processed the continuous data of a dense dataset of one year of OBS data along the BTFS, focusing on creating a high-resolution seismicity catalog. Rigorous processing from seismicity with different P- and S- phase pickers results in high-resolution seismicity distributions, which allow the mapping of individual fault strands of a transform, similar to strike-slip faults onshore. In particular, combining different phase pickers and using pick-time differences as proxies for pick uncertainty further improved the locations. The local seismicity indicates substantial along-strike variations and reveals the internal structure of the pull-apart basins and seismic and aseismic regions. Besides the main faults, seismicity occurs along basins and within, on aborted fracture zones, and in one off-fault cluster showing a swarm-like activity in 2008, reflecting the immature western BTFS complexity as its reorganization is ongoing.

We determined repeating events using the locations from the relative seismic catalog to infer the slip of the fault system in time. We find that on average the repeaters release 21 cm slip, which exceeds the slip rate of the overall fault system by ~4 times, indicating a stick-slip behavior of the fault and suggesting repeat times of ~4 years. Although it is long known that Transform faults are mostly aseismic, the repeaters show that this strain release is episodic, likely related to very small patches of the fault, which alternate between fully locked and fully creeping behavior in a stick-slip behavior. Repeaters at mantle depths along the Western Blanco ridge and an absence of crustal events indicate a fully locked fault segment. The other segments of the BTFS are at least partly creeping. Local earthquake tomography shows elevated vp/vs values for most of the Blanco transform fault system, likely related to significant serpentinization. Overall, the high number of 54 OBS allows us to map the structure of an oceanic plate boundary system. Due to the advances in machine learning techniques, we could apply methods such as repeaters or local earthquake tomography due to the availability of high-quality phase onsets, which help to resolve the dynamic nature of the Earth's crust.

**Acknowledgments**

The authors thank all individuals and organizations who contributed to data collection and supported the Blanco Transform OBS experiment (Nábělek and Braunmiller, 2012). IG and YR benefited from funding by the European Research Council (ERC) under the framework programme Horizon Europe of the European Union (ERC-TRANSFORMERS-101096190). YR was partly supported by the China Scholarship Council (Grant 201904910466) and the 2022 Beacon Prize sponsored by Zhihu Incorporation. We acknowledge funding for publication costs by the Projekt DEAL.

**Data availability**

The continuous seismological data (Kuna et al., 2019; Nábělek and Braunmiller, 2012) were provided by the Ocean Bottom Seismic Instrument Center (https://obsic.whoi.edu/) and are accessible from the IRIS (www.iris.edu) using the network code X9 for 2012–2013 (https://doi.org/10.7914/SN/X9_2012). The bathymetry was downloaded from the Global Multi-Resolution Topography database (GMRT, www.gmrt.org, Ryan et al., 2009). The ISC-GEM catalogue (Di Giacomo et al., 2018; Storchak et al., 2015, 2013) is available at www.isc.ac.uk. Figures were made with GMT 4.0 (Wessel et al., 2013). The relative and absolute earthquake catalogue and the lists with repeaters can be accessed at (XXUPLOAD-TO-ZENODO-OR-PANGEA-for-final-verson-XX).

*Journal of Geophysical Research: Solid Earth*

Supporting Information for

**Seismotectonics and Slip Behavior of a Submarine Plate Boundary Fault from Seismicity Repeaters and Tomography using a high-resolution earthquake catalog from machine learning**

D. Lange[1], Y. Ren[1], and I. Grevemeyer[1]

[1] GEOMAR Helmholtz Centre for Ocean Research Kiel, Wischhofstr. 1-3, 24148 Kiel, Germany

**Contents of this file**

    Figures   S1 to S12
    Tables    S1
    Movies

**Additional Supporting Information (Files uploaded separately)**

    Captions for Movies S1 to S3

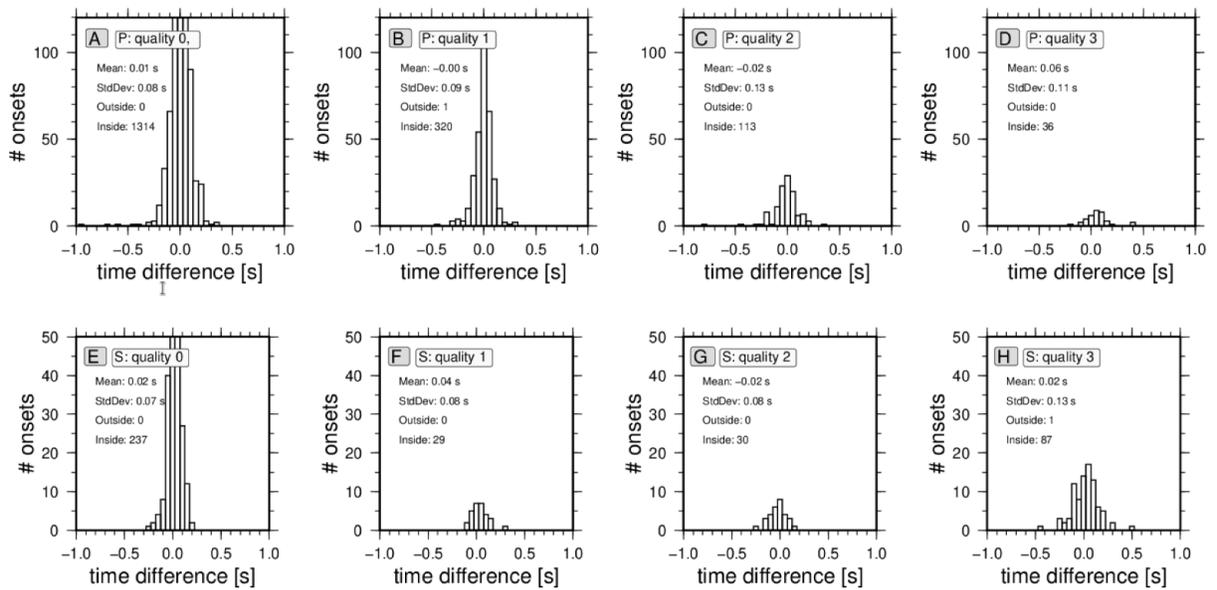

**Figure S1:** Histograms showing the arrival time differences between automated and manual picks from the final (absolute) catalog for the different weights from the human operator. Panels a,b,c, and d show the time-difference for P phases and panels e,f,g, and h indicate the histograms for the S-phases. The manually picked phases are from (Ren et al., 2023). Please note that these manual picks were part of one of the datasets used to train the OBS picker (Bornstein et al., 2024), so the plots might indicate an optimistic behavior of the picking engine compared to picks that were not used during the training.

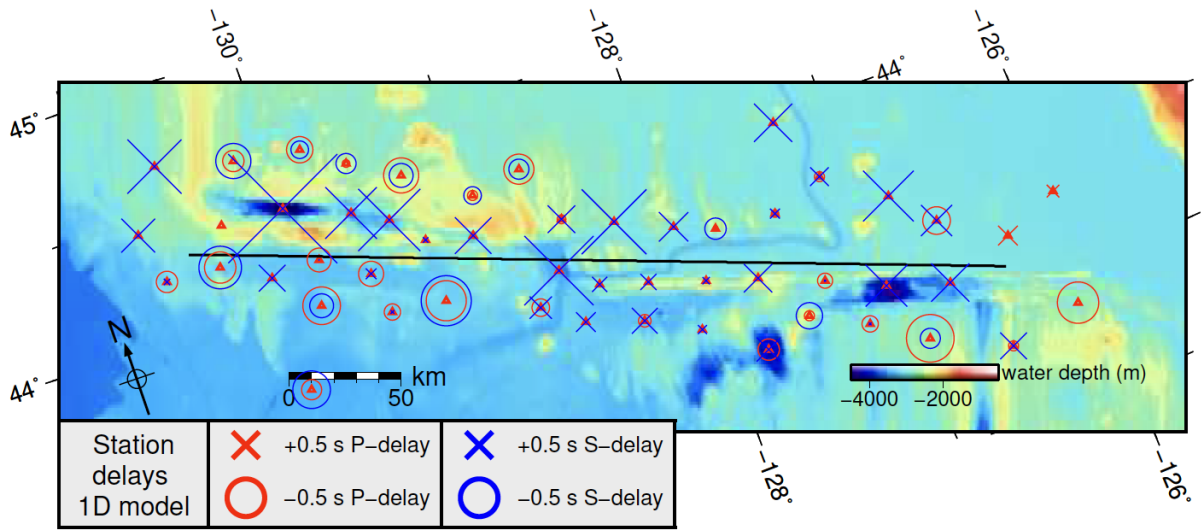

**Figure S2:** Stations delay times for the absolute catalog.

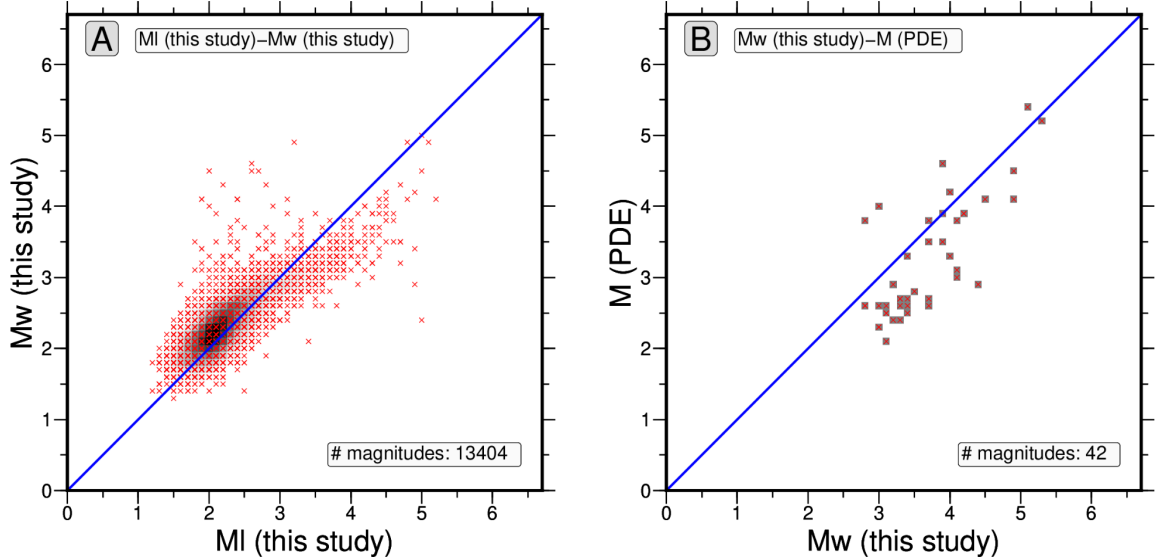

**Figure S3:** Properties of the magnitudes estimated from the local network data. (a) Ml versus Mw from the local seismicity catalogue. The grey colors in the background indicate the number of magnitude pairs. The blue line has a slope of 1. (b) Mw plotted versus magnitudes from the NEIC catalog.

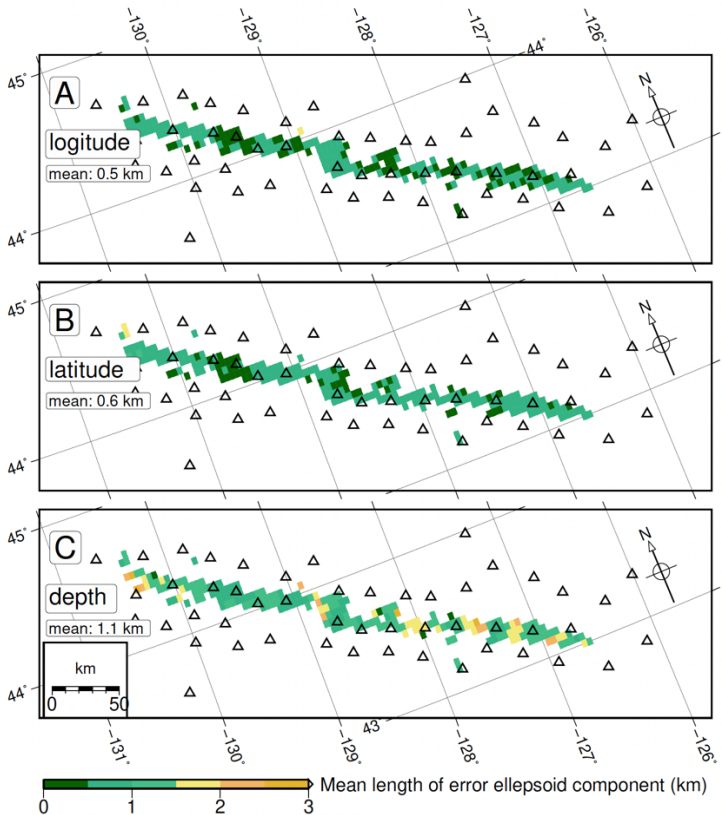

**Figure S4:** Mean uncertainties based on error ellipsoids (68% confidence) from NonLinLoc for 0.25×0.25° tiles for the absolute NonLinLoc catalog. (a) Mean longitude uncertainty, (b) latitude uncertainty, (c) depth uncertainty. Stations are shown with black triangles.

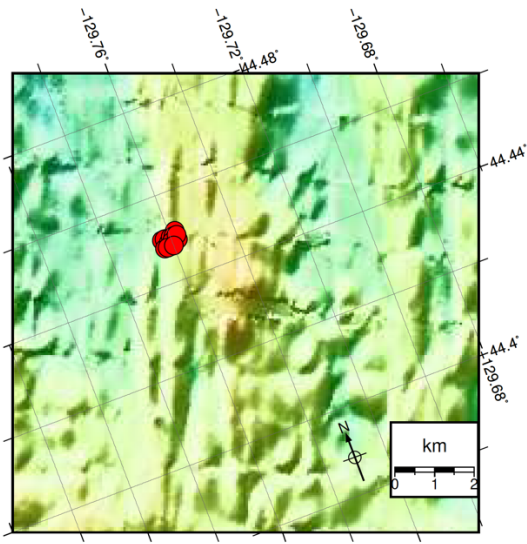

**Figure S5:** Closeup view of seismicity from the relative catalog for the off-axis cluster (15 Events). The 15 events could be located, all within 1 km (500m) distance, suggesting a conservative horizontal uncertainty of ~1 km (500 m), respectively. The location of the cluster is marked in Figure 1 with a red arrow.

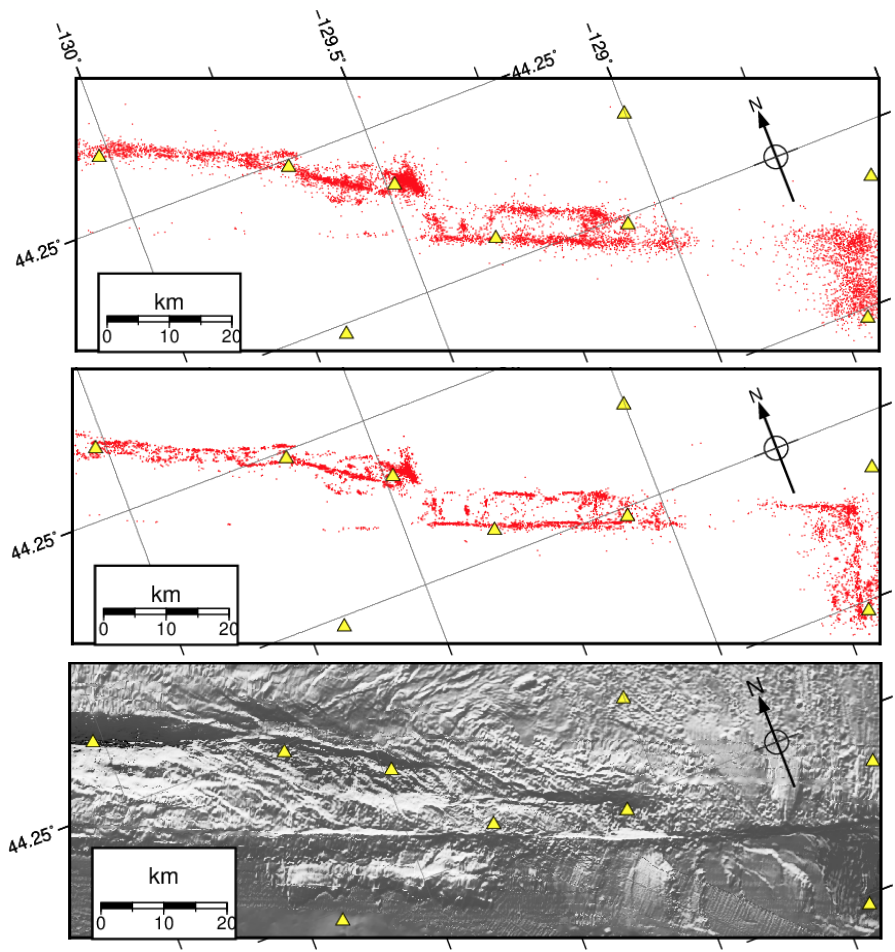

**Figure S6:** Comparison of the absolute seismicity catalog (upper panel) and the relative catalog (middle panel). The bathymetry is plotted in the lower panel. The closeup shows the region of the East Blanco Depression, Surveyor Depression and parts of the seismicity related to the Cascadia Depression. OBS stations are shown with yellow triangles.

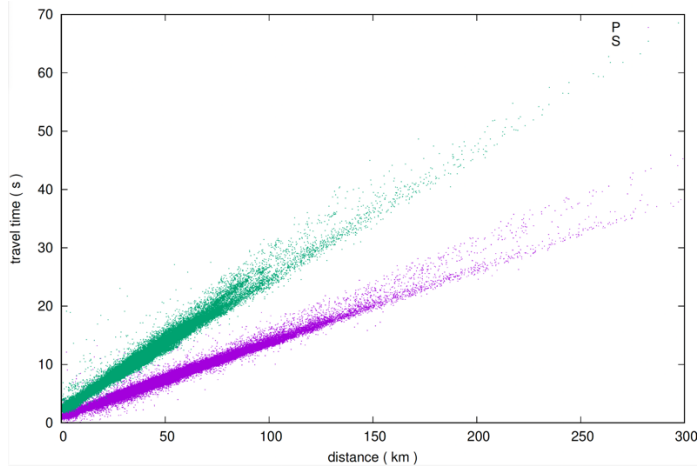

**Figure S7:** Travel time plotted against epicentral distance.

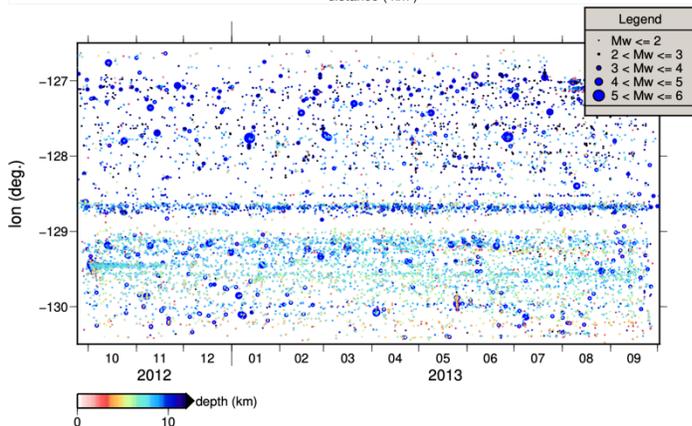

**Figure S8:** Longitudes of seismicity from the relative catalog plotted against time and color-coded with depth. The size of the circles represents the moment magnitude.

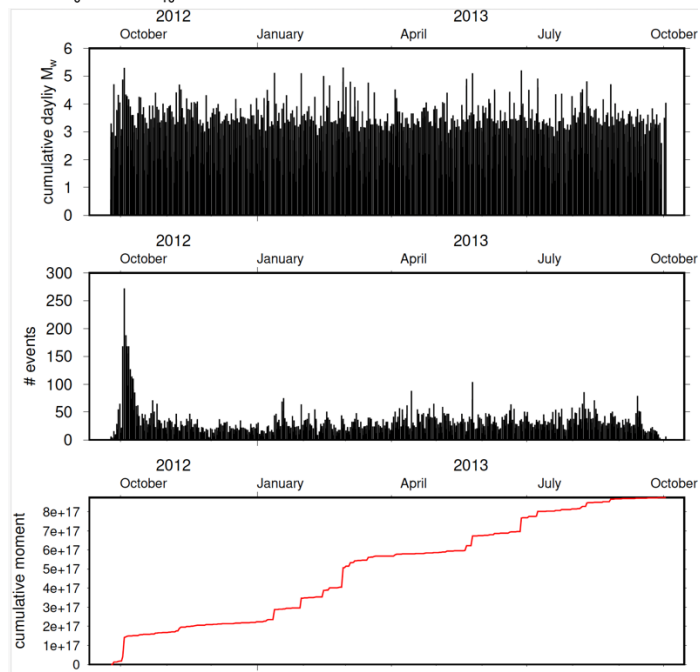

**Figure S9:** Temporal properties of the absolute seismicity catalog. (a) cumulative daily Mw (estimated by adding daily seismic moments (M0) and converting this back to moment magnitude (Mw) shown against time. (b) Number of events plotted vs. time. (c) Cumulative moment release over time.

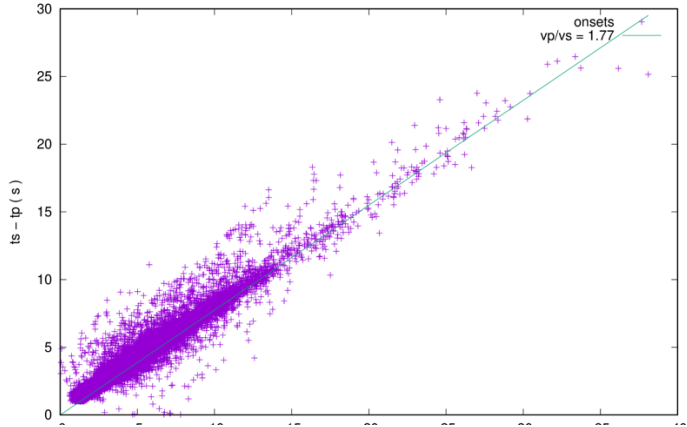

**Figure S10:** Wadati plot for the final absolute catalog. The green line indicates the slope of and vp/vs ratio of 1.77.

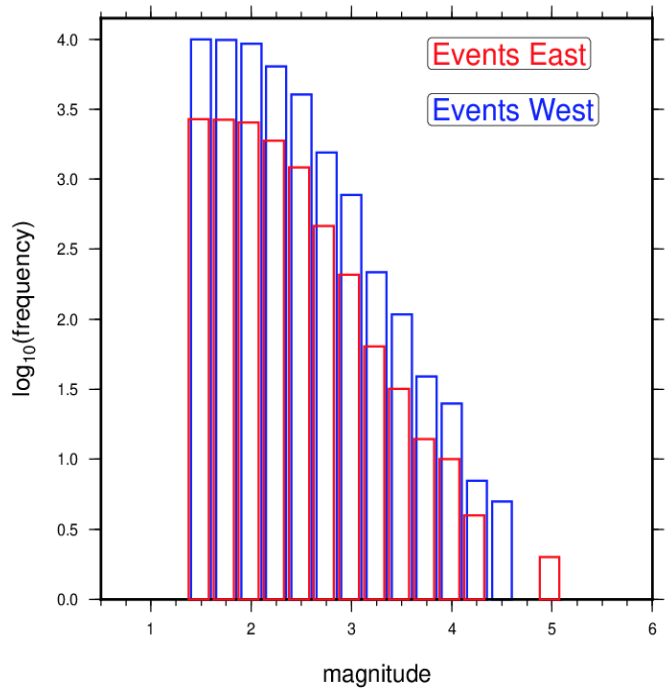

**Figure S11:** Histogram of Ml for the eastern and western BTFS. The catalog was split between East and West at 128.5° longitude.

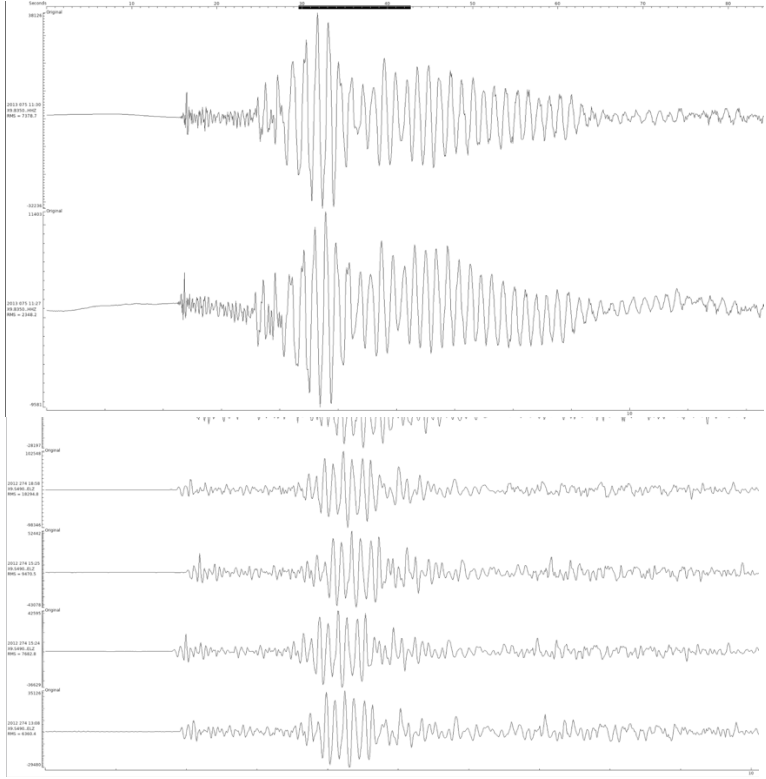

**Figure S12:** Data examples for 80 s long seismograms showing a repeater pair (upper panel) and a group of five repeaters (bottom).

**Table S1:** Properties for the different earthquake catalogs

| RMS | # events | Uncertainty | | | Catalogue | Remarks |
|---|---|---|---|---|---|---|
| | | longitude | longitude | depth | name | |
| (s) | (unitless) | (km) | (km) | (km) | | |
| 0.246 | 12731 | - | - | - | HYPODD | Relative locations from HYPODD |
| 0.257 | 13503 | 0.56 | 0.60 | 1.15 | ABS | Absolute locations from NonLinLoc, Picks from BluePhaseNet und BlueEQTransformer |
| 0.292 | 13502 | 0.45 | 0.50 | 0.90 | ABSOLUTE_ EQUAL_WEI GHTS | Absolute locations from NonLinLoc without weights |
| 0.185 | 12163 | 0.64 | 0.67 | 1.43 | PB-PHA-EQT | Catalogue with picks from BluePhaseNet und BlueEQTransformer (Bornstein et al., 2024) |
| 0.209 | 12163 | 0.56 | 0.59 | 1.23 | PB-PHA-EQT_EQUAL _WEIGHTS | Catalogue with picks from BluePhaseNet und BlueEQTransformer (Bornstein et al., 2024) without weights |
| 0.186 | 8525 | 0.74 | 0.71 | 1.38 | PB-EQT | Catalogue with picks from BlueEQTransformer (Bornstein et al., 2024) |
| 0.173 | 8539 | 0.77 | 0.75 | 1.53 | PB-PHA | Catalogue with picks from BluePhaseNet (Bornstein et al., 2024) |
| 0.462 | 8864 | 0.50 | 0.57 | 0.90 | EQT-LAND | Catalogue with picks from EQTransformer (Mousavi et al., 2020) |

**Movie S1:** Comparison of the catalogs based on the different pickers and different weights. The catalog labels are explained in Table ST1.

**Movie S2:** Comparison of the close-up views of catalogs based on the different pickers and weights. The catalog labels are explained in Table ST1.

**Movie S3:** Animation of daily seismicity of the relative earthquake catalog.